\documentclass[conference]{IEEEtran}
\makeatletter
\def\ps@headings{%
\def\@oddhead{\mbox{}\scriptsize\rightmark \hfil \thepage}%
\def\@evenhead{\scriptsize\thepage \hfil \leftmark\mbox{}}%
\def\@oddfoot{}%
\def\@evenfoot{}}
\makeatother 

\usepackage{framed}
\usepackage{balance}\balance
\usepackage{multicol}
\usepackage[normalsize]{caption}
\usepackage{amssymb}
\usepackage{cite}
\usepackage{times}
\usepackage{amsmath}
\usepackage{graphicx}
\usepackage{graphics}
\usepackage{epsfig}
\usepackage{latexsym}
\usepackage{amsfonts}
\usepackage{amssymb}
\usepackage{paralist}
\usepackage{xspace}
\usepackage{mathrsfs}
\usepackage{psfrag}
\usepackage{esint}
\usepackage{color}
\usepackage[noend]{algorithmic}
\usepackage[vlined,boxed,commentsnumbered]{algorithm2e}
\usepackage{subfigure}
\usepackage{multirow}
\usepackage{eso-pic}
\usepackage{url}%
\balance
\providecommand{\U}[1]{\protect\rule{.1in}{.1in}}
\def\ie{\textit{i.e.}\xspace}
\def\etal{\textit{et al.}\xspace}

\def\eg{\textit{e.g.}\xspace}
\def\st{\xspace\textbf{s.t.}\xspace}

\def\maxlink{{R}}

\def\sinrbnd{{\sigma}}
\def\noise{{\xi}}

\newtheorem{theorem}{\textbf{Theorem}}

\newtheorem{corollary}{Corollary}
\newtheorem{definition}{Definition}
\newtheorem{lemma}{\textbf{Lemma}}

\begin{document}

\title{ML$^2$S:
 \underline{M}inimum \underline{L}ength \underline{L}ink \underline{S}cheduling Under Physical Interference Model
\thanks{}
}
\author{\IEEEauthorblockN{Xiaohua Xu}
\IEEEauthorblockA{Department of Computer Science\\
Illinois Institute of Technology\\
Chicago, IL 60616\\
Email: xxu23@iit.edu}\and \IEEEauthorblockN{Jiannong Cao}
\IEEEauthorblockA{Department of Computing\\
Hong Kong Polytechnic University\\
Kowloon, Hong Kong\\ Email: csjcao@comp.polyu.edu.hk} \and
\IEEEauthorblockN{Xiang-Yang Li}
\IEEEauthorblockA{Department of Computer Science\\
Illinois Institute of Technology\\
Chicago, IL 60616\\
Email: xli@cs.iit.edu}}

\maketitle

\begin{abstract}
We study a fundamental problem called Minimum Length Link Scheduling
(ML$^2$S) which is crucial to the efficient operations of wireless networks.
Given a set of communication links of \emph{arbitrary} length spread and assume
 each link has one unit of traffic demand in wireless networks,
 the problem ML$^2$S seeks a schedule
for all links (to satisfy all demands)  of minimum number of
time-slots  such that the links assigned to the same time-slot do
not conflict with each other under the physical interference model.
In this paper, we will explore this problem under three
 important transmission power control settings:
linear power control, uniform power control and arbitrary power
control.  We design a suite of new and novel scheduling algorithms
and conduct explicit complexity analysis to demonstrate their
efficiency. Our algorithms can account for the presence of
background noises in wireless networks. We also investigate the
fractional case of the problem ML$^2$S
 where each link has a fractional demand.
 We propose an efficient greedy algorithm of the approximation ratio at most $(K+1)^{2}%
\omega$.

\end{abstract}

\section{Introduction}
\label{sec:introduction}
Wireless link scheduling plays a critical role
  in wireless networks
 and has been extensively studied in the literature
 \cite{goussevskaia2007cgs,roger09info,wan2009maximum}.
The task of link scheduling (or medium access control) is
challenging due to the simultaneous presence of two characteristics:
wireless interferences among concurrent transmissions, and the need
for practical distributed implementation with small communication
overhead and time complexity. The task will be more challenging if
there is a request for the performance guarantee of a link
scheduling algorithm. It is well known that a number of scheduling
problems (\textit{e.g.} \xspace, maximum throughput scheduling)
become NP-hard  when considering wireless interferences, while their
counter-parts are solvable in polynomial time for wired networks.
Thus, the scheduling algorithms for wireless networks often rely on
heuristics that approximately optimize the throughput.

As we know, the signal interferences cast significant effect on
the data throughput (or capacity region)
 that any scheduling algorithms can
 achieve.
Thus, we need to model and take care of wireless interferences
carefully to ensure the correctness of algorithm design. Wireless
link scheduling under various graph-based interference models (\eg,
the protocol interference model, CTS / RTS model, $K$-hop
interference model)
 have been extensively studied in the literature
\cite{joo2009understanding,wan2009multiflows}.
However, there are not many positive results on
 a series of problems related to link scheduling
under the physical interference
model\cite{kompella2010optimal,luo2010engineering}, These problems
receive great research interests
 recently in network
community for its accurate capture of signal interferences.
Under the physical interference model, we will model a successful transmission as follows:
 a signal is received
successfully if the Signal to Interference-plus-Noise Ratio (SINR) is above
 a certain threshold.
This threshold depends on hardware and coding method.

By using this practical interference model, a successful
transmission also accounts for interferences generated by
transmitters located far away, this fact greatly differs that in
graph-based interference
 models.
Therefore, traditional
algorithms for protocol interference model, RTS/CTS model \etal
 cannot be directly applied here.
We will focus on the physical interference model and address a
fundamental problem for wireless link scheduling called Minimum
Length Link Scheduling (ML$^2$S) or Shortest Link Scheduling. Given
is a set of links of arbitrary length diversity (or spread), assume
each link has one unit demand,
 the objective is to schedule all links (or to satisfy all demands)
 within a minimum number of time-slots; at the same time,
 the links scheduled in the same time-slot do not conflict
 with each other under the physical interference model.
This problem has several variations, one important variation is to
consider the problem under different transmission power control
settings. We will focus on three general power assignment settings:
 linear power control, uniform power control, and arbitrary power
 control.

\textbf{Linear Transmission Power Control\cite{xu2010maximum}:} Each
link $l$ (or its corresponding sender) is assigned with a
transmission power $c\cdot\Vert l\Vert^{\beta}$, where $c$ is a
constant and $0<\beta\leq\kappa$ is a constant. As a long
communication link requires more transmission power,
 this power control setting is usually energy-efficient.
Note that linear power assignment is a representative of the
\emph{oblivious} power assignment family where the transmission
power of each communication link only depends on its length.

\textbf{Uniform Transmission Power Control
\cite{goussevskaia2007cgs,roger09info}:} Each link is assigned with
 an identical transmission power.
The uniform power setting has been widely adopted in the literature
while it is notoriously hard to achieve a
 constant approximation for several variants of the link scheduling
 problems under this setting;
 for instance, to find constant approximations for both problems ML$^2$S and \emph{Maximum Weight Independent Set of Links (MWIL)} remain open.
Given a set of communication links, the problem MWIL
 seeks a subset of maximum weight, such that all links in this subset
 can transmit concurrently.

\textbf{Arbitrary Transmission Power Control:} Each link is assigned
with
 a transmission power of arbitrary value.
Clearly, this assignment is the most general.

Our main contribution is as follows. We study the problem ML$^2$S
under the physical interference model with
 different transmission power assignments.
 Under the linear power
 control setting, we are the first to propose algorithm design for the
 problem ML$^2$S  and prove the NP-hardness of this problem.
We use a partition strategy to find multiple sets of
 well-separated links and combine them to form a schedule. We
 prove that the solution returned by our algorithm satisfies the
  wireless interference constraint and can achieve constant
 approximation bound compared to the optimum one.
Under the uniform power
 control setting, we propose two algorithms that can achieve an approximation bound of $\min\{O(\log
\max\{f(l_i)\}), O(\log n)\}$ for the problem ML$^2$S. We also
present an efficient linear programming-based method for this
problem under the arbitrary power control setting. On the other
hand, we investigate the fractional case of the problem ML$^2$S
 where each link has a fractional demand.
We propose an efficient greedy algorithm and show that its approximation ratio is at most $(K+1)^{2}%
\omega$.

The rest of the paper is organized as follows. Section
\ref{sec:model} formulates the problem ML$^2$S and presents the complexity analysis.
Section \ref{sec:linear}, \ref{sec:uniform} and \ref{sec:arbitrary} are devoted to the algorithm design for
 the problem ML$^2$S under three transmission power control settings respectively.
Section \ref{sec:frac} deals with the fractional case.
Section \ref{sec:review} outlines the related work.
Finally, Section \ref{sec:conclusion} concludes the
paper.

\section{Network model}
\label{sec:model}

Assume we are given a set of links $A$ where the networking nodes $V$
 (\ie, sending and receiving nodes of all links) lie in plane,
 each node has a transmission power upper-bounded by $P_{\max}$ (or simply $P$).
Denote the Euclidean distance between a pair of
nodes $u$ and $v$ by $\left\Vert uv\right\Vert $.

Let $r$ be the smallest
distance between any pair of nodes in $V$. The path-loss over
a distance $l$ is $\eta l^{-\kappa}$, where $\kappa$ is
\emph{path-loss exponent} (a constant greater than $2$ and $5$
depending on the wireless environment), and $\eta$ is the
\emph{reference loss factor}. Since the path-loss factor over the
distance $r$ is less than one, we have $\eta<r^{\kappa}$.

We will consider three power assignments.
In the uniform power assignment, each link is assigned with the same
transmission power \cite{goussevskaia2007cgs,roger09info}.
In arbitrary power assignment, the power of each link can be arbitrary.
In an \emph{oblivious power control setting} \cite{fanghnel2009oblivious}, a node $u$ transmits to another
node $v$ always at a power depending on the length of the link.
In this work, we will assume the power for link $\overrightarrow{uv}$ is $c\left\Vert uv\right\Vert ^{\beta}$
for some constants $c>0$ and $0<\beta\leq\kappa$.
When $\beta=\kappa$, this implies a \emph{linear power assignment}.
This assumption
implicitly imposes an upper bound on the distance between a pair of
nodes which directly communicate with each other: For $u$ to be
able to directly communicate with $v$, we must have $c\left\Vert
uv\right\Vert ^{\beta }\leq P$ and hence $\left\Vert uv\right\Vert
\leq\left(  P/c\right) ^{1/\beta}$.

Let $\xi$ be the noise power, and $\sigma$ be the \emph{signal to
interference and noise ratio }(\emph{SINR}) threshold for successful
reception. Then, in the absence of interference, the transmission by
a node $u$ can be successfully received by another node $v$ if and
only if $ \tfrac{c\left\Vert uv\right\Vert
^{\beta}\cdot\eta\left\Vert uv\right\Vert ^{-\kappa}}{\xi}\geq\sigma
$
which is equivalent to $\left\Vert uv\right\Vert ^{\kappa-\beta}\leq
\frac{c\eta}{\sigma\xi}$.
Note that when $\left\Vert uv\right\Vert ^{\kappa-\beta}=
\frac{c\eta}{\sigma\xi}$, link $uv$ can only transmit alone
 since any other link will conflict with $uv$.
Thus we can disregard these links in $A$ and assume that
\[
\left\Vert uv\right\Vert ^{\kappa-\beta}<
\frac{c\eta}{\sigma\xi}
\]
Therefore, the set $A$ of
communication links consists of all pairs $\left(  u,v\right) $ of distinct
nodes satisfying that $\left\Vert uv\right\Vert ^{\beta}\leq P/c$ and
$\left\Vert uv\right\Vert ^{\kappa-\beta}<\frac{c\eta}{\sigma\xi}$. Let $R$
be the maximum length of the links in $A$. Then,
\[
R^{\kappa}=R^{\beta}\cdot R^{\kappa-\beta}<\frac{P}{c}\cdot\frac{c\eta
}{\sigma\xi}=\frac{P\eta}{\sigma\xi}<\frac{P}{\sigma\xi}r^{\kappa}.
\]
Hence,
\[
\frac{R}{r}<\left(  \frac{P}{\sigma\xi}\right)  ^{1/\kappa}.
\]

\bigskip

For a set of links ${\cal L}$, the \emph{length diversity} (or link
spread) is defined as the $\log\frac{\max\{\|l_i\|:l_i\in{\cal
L}\}}{\min\{\|l_i\|:l_i\in{\cal L}\}}$. Given an input links of
arbitrary length spread,
 a set $I$ of links in $A$ is said to be \emph{independent} if and only if all
links in $I$ \ can transmit successfully at the same time under the
physical interference model with oblivious
power assignment, \ie,  the SINR\ of each link in $I$ is above $\sigma$. We
denote by $\mathcal{I}$  the collection of independent sets of links in $A$.
\ Given a set of input links $A$, assume each link has one unit demand, the problem
\textbf{Minimum Length Link Scheduling} (\textbf{ML$^2$S}) or Shortest
Link Scheduling
seeks a schedule (partitioning the input links into multiple disjoint
subsets) of minimum length
 such that each subset of links are {independent}.

A fractional link schedule
is a set
\[S =\{(I_j , \gamma_j) : 1 \le j \le k\}\]
with $I_j\in {\cal I}$, and $\gamma_j \in \mathbb{R}^+$ for each $1
\le j \le k$.
The value $\sum^k_{j=1} \gamma_j$ is referred to as the length of the schedule $S$. We define the link load function $c_S \in
\mathbb{R}^A_+$ supported by the fractional schedule $S$ as
\[c_S (e) = \sum^k_{j=1} \gamma_j\cdot |I_j \cap \{e\}|, \forall e\in
A\]

Suppose we are give a set of links, each link $e$ is associated with a
fractional demand $d(e)$,
the problem \textbf{Minimum Length Fractional Link Scheduling} (\textbf{FML$^2$S}) or Shortest
 Fractional Link Scheduling seeks a fractional link schedule (each subset of links is an \emph{independent set of links}),
 of shortest length, such that for each link $e$, the demand
 $d(e)=c_S(e)$.
Please refer to \cite{wan2009multiflows} for a complete and detailed
definition on the fractional link schedule.

\bigskip

\textbf{Complexity of the problem ML$^2$S:}
We will prove that the problem ML$^2$S is NP-hard.
The  Partition problem has been proved to be NP-hard \cite{karp2010reducibility},
 thus we only need to reduce the Partition problem to ML$^2$S.
\begin{theorem}\label{the:np}
The Partition problem is reducible to the problem ML$^2$S in polynomial
time (The proof is available in the appendix).
\end{theorem}

\section{Linear Power Control}
\label{sec:linear}

In this section, we propose a distributed constant-approximation algorithm for
the ML$^2$S problem under the linear power control model.

\begin{figure}[b]\label{fig}
\begin{center}\scalebox{0.35}{\input{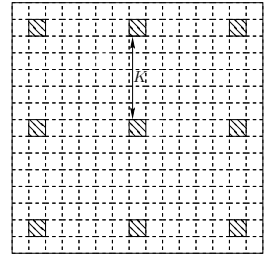}}\caption{A
partition of the plane into cells, each large-block consists of $K\times
K$ cells.}
\label{fig:cellpar_mwisl}
\end{center}
\end{figure}

Based on the definition of physical interference model,
 if we select a subset of geographically well-seperated
 links and let only these links transmit simultaneously,
 we can ensure that the interference on each link is bounded.
Using this property, we can schedule links in ${A}$ based
 on a partition scheme of the plane (Fig.~$1$)

For any link $l_i\in {A}$,
 we know $\|l_i\| \le \maxlink$.
Let $\ell=R/\sqrt{2}$. The vertical lines $x=i\cdot\ell$ for
$i\in\mathbb{Z}$ and horizontal lines $y=j\cdot\ell$ for
$j\in\mathbb{Z}$ partition the planes into half-open, half-closed
grids of side length $\ell$ (here $\mathbb{Z}$ represents the
integer set):
\[
\left\{  \lbrack i\ell,\left(  i+1\right)  \ell)\times\lbrack j\ell,\left(
j+1\right)  \ell):i,j\in\mathbb{Z}\right\}  .
\]

Based on a partition of the plane, we define a large-block as a
square which consists of $K\times K$ cells (grids). Here we let
\begin{equation}\label{equ:K}
 K=\lceil
\sqrt{2}\left({(4\tau)}^{-1}\left({{\sigma}^{-1}-\xi{(c\eta)}^{-1}R^{\kappa-\beta}}\right)\right)^{-1/\kappa}+\sqrt{2}\rceil
\end{equation}
Generally, $K$'s value depends on $R$, we can see that when
$\beta=\kappa$ (which is exactly the linear power assignment),\\ we
have,
\[
 K=\lceil
 \sqrt{2}\left({(4\tau)}^{-1}\left({{\sigma}^{-1}-\xi{(c\eta)}^{-1}}\right)\right)^{-1/\kappa}+\sqrt{2}\rceil
\]
which is a constant independent of $R$.
\begin{algorithm}[t]
\caption{Scheduling for ML$^2$S} \label{alg:par}
\SetKwInOut{Input}{Input}\SetKwInOut{Output}{Output}
\Input{Set of links $A=\{l_1, l_2, \cdots, l_n\}$.}
\Output{ Multiple independent sets ${\cal S}=\{S_1,S_2,\cdots,S_l\}$
 of links}
 Set $K$ according to Equation (\ref{equ:K})\;
\BlankLine
 $k \leftarrow 0$\;
\For{$r=0, \cdots , K$ and $s=0, \cdots, K$}{ \While{TRUE}{
  $k=k+1$\;
  \For{$i, j \in \mathbb{Z}$ and the cell $g_{i,j}$ contains links
    from ${A}$}{
    \If{$i \mod (K+1)=r$ and\\ $j\mod (K+1)=s$}{
    select one link whose sender lies within $g_{i, j}$;
    }
  }
  all the selected links form a set $S_k$\;
   \If{ $S_k=\emptyset$}{
    exit the while loop;
   }
   ${A}\leftarrow{A}\setminus S_k$\;
} } \Return {$S_1,S_2,\cdots,S_l$, here $l=k-1$.}
\end{algorithm}

For each time-slot, we let an independent set of links transmit,
 which is formed by picking at most one link from only the cell lying in
 the same relative location from every large-block.
We proceeds until all links have been picked already. The details of
our partition-based scheduling algorithm
 is shown in Algorithm~\ref{alg:par}.
The correctness of the algorithm follows from Lemma
\ref{l_sufficient}. Note that:
\begin{compactenum}
\item the output by Algorithm~\ref{alg:par} (the total number of time-slots needed)
 depends on the maximum number of links located in a large-blocks.
\item we do not keep any monotone ordering on the cardinality of link set scheduled in
each time-slot. In other words, for $i$-th time-slot, we may
schedule more links than the
 $(i-1)$-th time-slot ($i\in \mathbb{N}$).
\end{compactenum}

We first verify the correctness of our algorithm:
 each subset of links outputted by the proposed algorithm are
 independent (Lemma \ref{l_sufficient}).

For all $i,j\in\mathbb{Z}$, we denote $A_{ij}$ to be the set of
links in $A$ whose senders lie in the grid
\[
\lbrack i\ell,\left(  i+1\right)  \ell)\times\lbrack j\ell,\left(  j+1\right)
\ell).
\]

\begin{lemma}
\label{l_sufficient} Consider any two nonnegative integers $k_{1}$
and $k_{2}$ which are at most $K$. Suppose $I$ is a set of links
satisfying that for each  $i,j\in\mathbb{Z}$,  $\left\vert I\cap
A_{ij}\right\vert \leq1$ if
$i\operatorname{mod}\left(  K+1\right)  =k_{1}$ and $j\operatorname{mod}%
\left(  K+1\right)  =k_{2}$ and $\left\vert I\cap A_{ij}\right\vert
=0$ otherwise.  Then, $I$ is independent (The proof is available in
the appendix).
\end{lemma}

We then calculate the approximation ratio of the proposed algorithm.
The main idea is as follows: we first estimate  the number of
time-slots needed for Algorithm~\ref{alg:par} and derive an
upper-bound. Then, we will compute a lower-bound on the number of
 time-slots required for any algorithm to schedule links lying inside
 a single
 large-block, thus a lower-bound for the length of any schedule to
 transmit all input links.
By comparing these upper-bound and lower-bound,
 we can compute the approximation ratio of the proposed algorithm.

Based on the partition scheme (Fig.~\ref{fig:cellpar_mwisl}), we
assume the maximum number of senders from ${A}$ lying inside any
small cell is $B$. Then, we have the following lemma on the
upper-bound of the number of time-slots needed for
Algorithm~\ref{alg:par}.
\begin{lemma}\label{lem:len}
The proposed algorithm costs at most $(K+1)^2 \cdot B$ time-slots.
\end{lemma}
\begin{proof}
As shown in line $3$ of Algorithm~\ref{alg:par}, there are at most
$(K+1)^2$ while loops. After each iteration of a while loop, the
links whose sender lies inside each cell will decrease by at least
one. Thus, the number of $B$ iterations for the while loop is at
most $B$. In total, Algorithm~\ref{alg:par} costs at most $(K+1)^2
\cdot B$ time-slots.
\end{proof}

Next, we calculate the lower-bound of any algorithm for ML$^2$S that
is the minimum possible length for any algorithm.

\bigskip

Let
\[
\omega=\left\lceil \frac{2^\kappa P}{\sigma^2\xi} +1\right\rceil .
\]
\begin{lemma}
\label{l_necessary} For any $I\in\mathcal{I}$ \ and any
$i,j\in\mathbb{Z}$, \ $\left\vert I\cap A_{ij}\right\vert
\leq\omega$.
\end{lemma}
\begin{proof}
Let $I_{ij}=I\cap A_{ij}$. Assume $a=\left(  u,v\right)$ be the
shortest link in $I_{ij}$, consider any link $a^{\prime}=\left(
u^{\prime},v^{\prime}\right)  $ in $I_{ij}$ other than $a$, the
distance between the sender $u^{\prime}$ and $v$ satisfies {\small\[
\left\Vert u^{\prime}v\right\Vert\le \left\Vert
u^{\prime}u\right\Vert+ \left\Vert uv\right\Vert \le \left\Vert
\sqrt{2}\ell\right\Vert+ \left\Vert uv\right\Vert\le 2R,\]}
The SINR at $a$ from all other links in $I_{ij}$ is at most%
\begin{align*}
& \frac%
{c\left\Vert a\right\Vert ^{\beta}\cdot\eta\left\Vert a\right\Vert
 ^{-{\kappa}}}{\sum_{a^{\prime}\in I_{ij}\setminus \{a\}}
{c\left\Vert a^{\prime}\right\Vert ^{\beta}\cdot\eta\left\Vert u^{\prime}v \right\Vert ^{-{\kappa}}}}\\&
\le \frac{\left\Vert a\right\Vert
 ^{-{\kappa}}}{ \sum_{a^{\prime}\in I_{ij}\setminus \{a\}} \left\Vert
 u^{\prime}v \right\Vert ^{-{\kappa}} }\\&
 \le \frac{r ^{-{\kappa}}}{ \left(\left\vert I_{ij}\right\vert-1 \right) \left(2R \right) ^{-{\kappa}} }\\
\end{align*}
Since $\frac{r ^{-{\kappa}}}{ \left(\left\vert I_{ij}\right\vert-1
  \right) \left(2R \right) ^{-{\kappa}} }\ge\sigma$, we have $\left\vert I_{ij}\right\vert \le
\frac{2^\kappa}{\sigma} \frac{R^\kappa}{r^\kappa}
  +1<\frac{2^\kappa}{\sigma} \frac{P}{\sigma\xi} +1
=\frac{2^\kappa P}{\sigma^2\xi} +1 $.
\end{proof}

\bigskip

Since there exists one cell that contains $B$ senders, by
Lemma~\ref{l_necessary}, this fact immediately implies a lower-bound
on the number of
 time-slots required for any algorithm to schedule links lying inside a
 large-block, and thus a lower-bound for any algorithm to schedule all input links.:

\begin{corollary}
Any valid schedule for ML$^2$S has a length at least
$\frac{B}{\omega}$.
\end{corollary}

\begin{theorem}
The approximation ratio of our algorithm for ML$^2$S is at most $(K+1)^{2}%
\omega$.
\end{theorem}
\begin{proof}
Since our algorithm has a length at most $(K+1)^2 \cdot B$, at the
same time, any valid schedule for ML$^2$S has a length at least
$\frac{B}{\omega}$, thus the approximation ratio of the proposed
algorithm can be bounded by
 $(K+1)^2 \cdot \omega$.
So, the theorem holds.
\end{proof}

\section{Uniform Power Control}
\label{sec:uniform} In this section, we design a scheduling for the
problem ML$^2$S
  under the uniform transmission power setting.

\subsection{First Algorithm}
For each link $l_i\in A$,
 we first introduce a concept of \emph{conflict
range factor} which is defined as
 \[f(l_i)=\frac{1}{1-\|l_i\|/R} \]
The conflict range factor can indicates how far should a set of
links
  of similar lengths be seperated to ensure concurrent transmissions.
Observe that the concept of \emph{relative interference} as defined
in
 \cite{wan2009maximum} is closely
 related the {conflict range factor}.

We then calculate the conflict range factor of each link and
 partition the links into $\log \max\{f(l_i): l_i\in A\}$ groups.
The first group consists of the links
 with the {conflict range factor} at most one.
The $i$-th ($2\le i\le \log \max\{f(l_i): l_i\in A\}$)
 group $G_i$ consists of the links
 with the {conflict range factor} lying in $[2^{i-1}, 2^{i}]$.
For each group $G_i$, we can use partition and shifting to find a
  schedule for all member links, which consists multiple scheduling
  sets.
We can output the union of all scheduling sets for all groups, as
our
 solution.
According to Lemma $1$ in \cite{wan2009maximum}, we can verify the
 correctness of our algorithm easily.

We next derive an approximation bound on the performance of our
algorithm.

\begin{theorem}
\label{tem:uniform} The number of scheduling set resulted from the
proposed algorithm is at most $\log \max\{f(l_i): l_i\in A\}$ times
the optimum solution.
\end{theorem}
\begin{proof}
Considering the group $G_i$ which costs the maximum number of
time-slots in our algorithm. Assume the number of time-slots to
transmit all links in $G_i$ by our algorithm is $k$. Thus our
algorithm costs at most $k\cdot \log \max\{f(l_i)\}$ time-slots.
Since all links in $G_i$ has similar {conflict range factor}, we can
verify that  any feasible schedule would take at least $O(k)$
time-slots to transmit all links in this group $G_i$. Thus the
approximation bound of our algorithm is $O(\log \max\{f(l_i)\})$.
\end{proof}

\subsection{Second Algorithm}
Our second algorithm consists in iteratively computing a maximum
independent set of links by using the algorithm in Table $1$ in
\cite{wan2009maximum}. We then transmit each independent set of
links  in a separate time-slot,
  and the remaining links are repeatedly used as input to the algorithm in Table $1$ in \cite{wan2009maximum}.
 The procedure continues until all links in $A$ have been scheduled.
The correctness of the obtained schedule has been proved as the
algorithm in Table $1$  in \cite{wan2009maximum} indeed outputs an
independent set of links. It is trivial show that the  approximation
bound of the second algorithm is $O(\log n)$.

We make a note that Goussevskaia\etal \cite{roger09info} made the
first
  effort on developing a $O(\log n)$ for the problem ML$^2$S under the
uniform transmission power control setting. However, as observed in
\cite{xu2009constant}, the claimed constant approximation bound and
its proof (Lemma 4.5 in \cite{roger09info}) are valid only when the
background noise is zero; thus their result becomes baseless.
Despite of this effort in \cite{roger09info}, the existence of a
$O(\log n)$-approximation
  algorithm for
 ML$^2$S under the
uniform transmission power control setting remains open.

To sum up, we can achieve an approximation bound of $\min\{O(\log
\max\{f(l_i)\}), O(\log n)\}$ for the problem ML$^2$S under the
uniform transmission power control setting.

\section{Arbitrary Power Control}
\label{sec:arbitrary} This section presents the link scheduling
under the physical interference model with arbitrary power setting.
We will borrow the idea of cover inequalities in
\cite{capone2011new}
 of solving the problem maximum independent set of links to our setting.
The technique of  cover inequalities  has found applications for
solving the $0$-$1$ knapsack problem \cite{atamtrk2005cover}.

Let the binary variable $x_{i, t}$ denote if the link
 $l_i$ is activated at the $t$-th time-slot or not.
Let the binary variable $y_t$ denote
 if there exists at least one link activated at the $t$-th
 time-slot or not.
Let $g_{ji}$ denote the path-loss from the sender of link $l_j$ to
the
 receiver of link $l_i$;
  thus $g_{ii}$ denotes the path-loss from the sender of link
 $l_i$ to its receiver of the same link.
Note that $P_i$ denotes the power of link $l_i$. If the link $x_i$
is active at the $t$-th time-slot,
 the SINR requirement can formulated by the following inequality.
\[P_i g_{ii} x_{i,t}  \ge  \sinrbnd (\sum P_j g_{ji}
 x_{j,t} + \noise)\]
Let $a_i=\frac{P_i\cdot g_{ii}}{\sinrbnd}-\noise$ and
$b_{ji}=P_j\cdot
 g_{ji}$, then this inequality can be written as a knapsack constraint:
  \[ \sum  b_{j,i} x_{j,t}\le a_i\].
Thus, for any link $l_i$ and any time-slot $t$, we have
\[a_i + M_{i}(1-x_{i,t}) \ge \sum b_{j,i} x_{j,t}\]
The inequality holds because when $x_{i,t} = 1$, this is exactly the
 SINR constraint; when $xij = 0$,
 this constraint can be satisfied for a large value of $M_i$.
To sum up, the integer linear programming is described as follows.
\begin{eqnarray}
\nonumber
\max \sum_{t=1}^n y_{t} \quad\quad \st \quad\quad\quad\quad\quad\quad\quad\quad\\
\begin{cases}
a_i + M_{i}(1-x_{i,t}) \ge \sum b_{j,i} x_{j,t}: \forall i, t\in\{1,2,\cdots,n\}; \\
 \sum_{t=1}^n x_{i, t}=1, \forall i, t\in\{1,2,\cdots,n\};\\
 y_t\ge x_{i, t}, \forall i, t\in\{1,2,\cdots,n\};\\
 x_{i, t}\in \{0,1\},  \forall i, t\in\{1,2,\cdots,n\};\\
 y_t \in \{0,1\}, \forall t\in\{1,2,\cdots,n\}.
\end{cases}
\end{eqnarray}

The second inequalities mean that each link must be activated in one
time-slot. The third inequalities reflect the constraint that if a
link is activated at some time-slot $t$, then $y_t=1$. In this
integer linear model, the big number $M_i$ would weaken the
continuous relaxation. Even worse, the coefficients of $x_{i,t}$ in
the  the knapsack constraint vary significantly
 in magnitude and would cause numerical difficulties when solving the problem.
Therefore, we reformulate the knapsack constraint by using cover
 inequality-type cutting planes.
A set $L$ of links is called a cover, if
 $\sum b_{j,i} \ge a_i$.
Due to the SINR constraint for link $l_i$, we can not transmit all
links in $L$ simultaneously. Then at most $|L|-1$ links in $C$ can
be active simultaneously and therefore
 the so-called \emph{cover inequality} holds:
\[\sum x_{j,t} \le |L|-x_{i,t} : \sum_{l_j\in L} b_{j,i} \ge a_i\]
Note that the inclusion of $x_{ij}$ in the right-hand side
 restricts the activation of links in $L$ to at most
 $|L|-1$, only if link $l_i$ itself is active.

\textbf{Strengthen Inequalities:} We still can not solve the LP in
its complete form, as the number
 of SINR cover inequalities grows exponentially.
Let $\{x_{i, t}\}$ denote an integer solution satisfying any subset
of
 SINR cover inequalities.
Verifying whether or not $\{x_{i,t}\}$ violates the
 SINR of any link $l_i$ with $x_{i,t} = 1$ is straightforward.
To get stronger inequalities, we generate (7) having a minimum
number of elements in the left-hand side, corresponding to so called
\emph{minimum cover} for knapsack problems. We minimize the number
of interfering nodes we pick before the sum exceeds $a_i$. Doing so
amounts to sorting the elements in $b_{ji}$
  in descending order, and following the
 sorted sequence until the accumulated sum goes above $a_i$.
Denoting the resulting index set by $K$, then $K$ is a minimal
cover. We strengthen it further by a lifting process, \ie,
subtracting additional $x$-variables in the right-hand side based on
the conflict
 constraints.
Then the general framework for solving the LP is described as
follows:
\begin{enumerate}
\item Solve the current LP, let $\{x_{i,t}\}$ be the optimal solution.
\item If $\{x_{i,t}\}$ is a feasible schedule, then we are done.
\item Otherwise, we find possible cover inequalities violated by
$\{x_{i,t}\}$ and add to the LP. Go to Step $(1)$.
\end{enumerate}

\section{Dealing with Fractional Demand}
\label{sec:frac} In this section, we study the problem FML$^2$S
under the physical interference model. We will focus on the linear
transmission power control setting. The proposed algorithm can be
adapted to some other power
 control settings as well.
\begin{algorithm}[t]
\label{alg:fractional}
\SetKwInOut{Input}{Input}\SetKwInOut{Output}{Output}
\Input{Set of links $A=\{l_1, l_2, \cdots, l_n\}$,\\
  the initial demand function $d\in \mathbb{R}^A_+$.}
\Output{A fractional link schedule:\ ${\cal S} =\{(I_j , \gamma_j) :
1 \le j \le k\}$.} \BlankLine
 $j \leftarrow 0$ \;
\While{there exist link(s) in $A$} { \For{$k_1=0, \cdots , K$ and
$k_2=0, \cdots, K$}{
  \For{$i, j \in \mathbb{Z}$ and the grid $g_{i, j}$ contains links from ${A}$}{
    \If{$i\equiv k_1 \mod (K+1)$ and $j\equiv k_2 \mod (K+1)$}{
    select one link whose sender lies within
 $g_{i, j}$  }
 }

Let all the selected links form a set $I_j$;

 \If{$I_j\neq \emptyset$}{
    $j++$;

Compute the minimum demand $d_{\min}$ from all links in $I_{j}$ as
follows:

$d_{\min} \leftarrow \infty$

\For{each link $e\in I_j$} {  \If{$d (e)< d_{\min}$}{
    $d_{\min}=d (e)$; }
}

Let $\gamma_j=d_{\min}$;

\For{each link $e\in I_j$} { Replace $d (e)$ by $d (e) -\gamma_j$;
  \If{$d (e)=0$}{
    Remove the link $e$ from $A$; }}
}} }
 \Return{A schedule ${\cal S}$ containing a subset of links in $A$.}
\caption{Fractional Link Scheduling.} \label{alg:fractional}
\end{algorithm}

We will construct a fractional link schedule ${\cal S}$ in an
iterative manner. At the beginning of each iteration,
 we denote the subset of links $e$ in $A$ with $d (e) > 0$ as $A^\prime$.
We use $A^\prime$ as the set of input links and
 apply a partition-based algorithm to select an independent set $I$ of
 links (may not be maximal) from $A^\prime$.
Let $\gamma = \min_{e\in I} d (e)$, and add $(I, \gamma)$ to ${\cal
S}$. We then update the demand function, by replacing $d (e)$ by $d
(e) -\gamma$ for each link $e$ in $I$.

We repeat this iteration until $d (e) =  0$ (the demand of this link
$e$ reaches zero) for every link $e\in A$. The details of the
partition-based scheduling algorithm are shown in
Algorithm~\ref{alg:fractional}.

Note that, after each iteration the number of links with positive
demand strictly decreases. Thus, Algorithm~\ref{alg:fractional} will
terminate in a finite number of iterations (bounded by $|A|$). The
correctness of Algorithm~\ref{alg:fractional} follows from Lemma
\ref{l_sufficient}. Next, we derive the approximation ratio of the
proposed scheduling algorithm.

To analyze the performance, we first upper-bound the length of the
schedule outputed by
 Algorithm~\ref{alg:fractional}.
 we then derive a lower bound on the length required for any schedule.
By comparing the two parts, we can derive the approximation ratio of
the proposed algorithm (Algorithm~\ref{alg:fractional}).

For each cell $g_{i,j}$, let ${A}_{i,j}$ be the set of all links
lying inside the cell $g_{i,j}$. Let $D_{i,j}$ be the sum of demands
for all links in ${A}_{i,j}$. Assume the maximum value of $D_{i,j}
:\forall i,j\in\mathbb{Z}$ is $D$ (\ie, the maximum total demand in
a single cell).

\begin{lemma}\label{lem:len}
The proposed algorithm has a length at most $(K+1)^2 \cdot D$.
\end{lemma}
\begin{proof}
Let $k_1\in\{0, \cdots , K\}$ and $k_2\in\{0, \cdots, K\}$. Consider
a number pair $(k_1,k_2)$, for each cell $g_{i, j}: i, j \in
\mathbb{Z}$ such that $i\equiv k_1 \mod (K+1)$ and $j\equiv k_2 \mod
 (K+1)$ hold, we select a link arbitrarily from the cell.
By Lemma \ref{l_sufficient}, we know that all the selected links can
transmit simultaneously. Using this property, we know that after
each time-slot, we can satisfy
 the demand from each cell $g_{i, j}$ by at least one.
Thus, to satisfy all demands for all cells d $g_{i, j}$ with
 $i\equiv k_1 \mod (K+1)$ and $j\equiv k_2 \mod
 (K+1)$,
we only need a fractional link schedule of length at most $D$,
 where $D$ the maximum total demand in a single cell

Now, we consider different  $k_1\in\{0, \cdots , K\}$ and
$k_2\in\{0, \cdots, K\}$, there are totally at most $(K+1)^2$
different pairs of $(k_1,k_2)$. For each pair, we only need a
fractional link schedule of length at most $D$. Thus, totally,
Algorithm~\ref{alg:par} outputs a schedule of length at most
$(K+1)^2 \cdot D$.
\end{proof}

Next, we show the lower-bound of any algorithm for FML$^2$S that is
the minimum length of any algorithm.

\bigskip

\begin{lemma}
Any valid schedule for FML$^2$S has a length at least
$\frac{D}{\omega}$.
\end{lemma}
\begin{proof}
By Lemma~\ref{l_necessary}, for any single cell, any valid schedule
can satisfy a total demand of at most $w$ in one time-slot. Since
there exists a cell which contains a total demand $D$, any
fractional link schedule has a length at least $\frac{D}{\omega}$ to
ensure interference-freeness.
\end{proof}

\begin{theorem}
The approximation ratio of our algorithm for FML$^2$S is at most $(K+1)^{2}%
\omega$.
\end{theorem}
\begin{proof}
Since our algorithm has a latency at most $(K+1)^2 \cdot D$, at the
same time, any valid schedule for FML$^2$S has a latency at least
$\frac{D}{\omega}$, thus the approximation ratio of the proposed
algorithm can be bounded by
 $(K+1)^2 \cdot \omega$.
So, the theorem holds.
\end{proof}

\section{Literature review}
\label{sec:review}

 Under the physical
 interference model, the problem of joint scheduling and power control has been well studied.
For instance, in \cite{elbatt2004jsa, cruz2003orl},
 optimization models and heuristics for this problem are proposed.
In \cite{gao2008tcm, moscibroda2006tcm}, topology control with SINR
 constraints is studied.
In \cite{moscibroda2006ccw},
 a power-assignment algorithm
 which schedules a strongly connected set of links
 in poly-logarithmic time is presented.
In \cite{chafekar2007cll}, the combined problem of routing and power
 control is addressed.

In \cite{goussevskaia2007cgs}, the scheduling problem without power
 control under physical interference model,
 where nodes are arbitrarily distributed in Euclidean space,
 has been shown to be NP-complete.
A greedy scheduling algorithm with
 approximation ratio of $O(n^{1-2/(\Psi(\alpha)+\epsilon)}(\log n)^{2})$,
 where $\Psi(\alpha)$ is a constant that depends on
 the path-loss exponent $\alpha$, is proposed in \cite{brar2006ces}.
Notice that this result can only hold when the nodes are
 distributed uniformly at random in a square of unit area.

In \cite{goussevskaia2007cgs}, the authors proposed
 an algorithm with a factor $O(g(L))$ approximation guarantee
 in arbitrary topologies, where $g(L)=\log \vartheta(L)$ is the
 diversity of the  network.
In \cite{chafekar2008aac},
 an algorithm with approximation guarantee of $O(\log \Delta)$ was
 proposed,
 where $\Delta$ is the ratio between
 the maximum and the minimum distances between nodes.
Obviously, it can be arbitrarily larger than $\vartheta(L)$.

Recently, Goussevskaia \etal \cite{roger09info} proposed a method
 for Maximum Independent Set of Links (MISL).
Unfortunately, as observed in Xu and Tang \cite{xu2009constant},
 their method \cite{roger09info} only works correctly in
 absence of the background noise.
Wan \etal \cite{wan2009maximum} resolved this issue by developing
 the first correct constant-approximation algorithm.
\cite{xu2010maximum} gave a constant-approximation algorithm for
 the problem of maximum weighted independent set of links under the oblivious power control setting.

Most Recently, Halldorsson \etal \cite{halldorsson2009wireless}
 presented a robustness result, showing that constant parameter
 and model changes will modify the minimum length link scheduling result
  only by a constant.
\cite{halldorsson2009adj} and \cite{fanghnel2009oblivious}
 studied the scheduling problem under power control respectively.
The minimum latency link  scheduling problem has been studied in
\cite{wan2010shortest}.

\section{Conclusions}
\label{sec:conclusion}

In this paper, we studied a fundamental problem called Minimum
Length Link Scheduling (ML$^2$S) which is crucial to the efficient
operations of wireless networks. We focused on the physical
interference model and the presence of background noises. We
considered this problem under three
 important transmission power control settings:
linear power control, uniform power control and arbitrary power
control. We designed a suite of novel scheduling algorithms and
conduct explicit complexity analysis to demonstrate the efficiency
of the proposed algorithms. In addition, we also investigated the
fractional case of the problem ML$^2$S and proposed an efficient
greedy algorithm with currently the best approximation ratio.

Some interesting questions are left for future research.
 The first one is to develop constant approximation
 scheduling algorithms for uniform transmission power and adjustable transmission power assignment settings.
The second one is to extend our algorithms to deal with a more
general
 path loss model.

\appendix
\subsection*{Proof of Theorem~\ref{the:np}}

We first introduce the \emph{Partition problem}:
\begin{definition}
Given a set ${\cal I}$
  of integers, whether we can divide ${\cal I}$ into two disjoint subsets
  ${\cal I}_1$ and ${\cal I}_2$, such that $I_1 \cup I_2 = {\cal I}$
  and the summations in each subset are equal?
\end{definition}

Given an instance of the Partition problem: ${\cal I} =
\{i_1,\cdots, i_n\}$ of integers with $\sum_{j=1}^{n} i_j =N$; we
will construct a ML$^2$S instance with $n + 2$ links
 $A =\{a_1,\cdots, a_{n+2}\}$, where $a_i=\overrightarrow{s_j,r_j}$.
Let the parameter $a=\frac{c/\sigma}{N/2}$. For each integer $i_j\in
{\cal I}$, we set the coordinate of  corresponding sender $s_j$ as:
$pos(s_i)=\left(\sqrt[\kappa]{\frac{P}{a i_j}}, 0\right)$;
 and the coordinate of receiver $r_i$ to be $ pos(s_i) + \left(b, 0\right)$,
 where $b =$
and $i_{\max}=\max\{i:i\in {\cal I}\}$. Finally, we place $r_{n+1}$
and $r_{n+2}$ at the center $(0, 0)$ and
 $s_{n+1}$, $s_{n+2}$ at $\left(0,\pm b\right)$ respectively.
For simplicity, let $\eta=1, \xi=0$.

In order to transmit successfully, the SINR constraint at the
intended receiver has to be satisfied. In Lemma~\ref{lem:np1},
 we will prove that the receivers $r_1,\cdots, r_n$ are close
enough to their respective senders to guarantee successful
transmissions, regardless of the number of other links scheduled
simultaneously.
\begin{lemma}\label{lem:np1}
Let $A_i = \{a_j: 1 \le j \le n+1 \ \& \ i\neq j\}$. It holds for
all $i\le n$ that the SINR exceeds $\sigma$ when the link $a_i$ is
scheduled concurrently with the set $A_i$.
\end{lemma}
\begin{proof}
Since the positions of the sender nodes $s_1,\cdots, s_n$ depend on
the values of $i_1,\cdots , i_n$, we can determine the minimum
distance between two sender nodes $s_j , s_k$: $\|s_j , s_k\| =
\|s_j , r_{n+1}\|- \|s_k, r_{n+1}\| =\sqrt[\kappa]{\frac{P}{a
i_j}}-\sqrt[\kappa]{\frac{P}{a i_k}}\ge
\sqrt[\kappa]{\frac{P}{a}}\cdot
\left(\frac{1}{\sqrt[\kappa]{i_{\max}-1}}-\frac{1}{\sqrt[\kappa]{i_{\max}}}\right)=f$.

Thus, any sender (including $s_{n+1}$) are located at least distance
$\|s_j , s_k\|-b$ away. By setting $b$ properly, we can ensure that
the successful transmission of $a_i$.
\end{proof}

Note that, any schedule to transmit all links needs at least two
time-slots, since $a_{n+1}$ and $a_{n+2}$ can never be scheduled
simultaneously, we then prove that:
\begin{lemma}
There exists a solution to the Partition problem if and only if
there exists a two-slot schedule for $A$.
\end{lemma}
\begin{proof}
\textbf{$\Rightarrow$:} Assume we know two subsets ${\cal I}_1,
{\cal I}_2 \subset {\cal I}$, whose elements sum up to $N/2$. To
construct a two-slot schedule, $\forall i_j \in {\cal I}_1$,
 we assign the link $a_j$ to the first time slot,
 along with $a_{n+1}$, and assign the remaining links to
the second time slot. We then check the correctness of our schedule.
The signal power that $r_{n+1}$ receives from $s_{n+1}$ is $c$ The
interference $r_{n+1}$ experiences from each sender $s_j$ is $ai_j$,
which results in the SINR at  $r_{n+1}$ of at least $\sigma$. This
fact, in combination with Lemma~\ref{lem:np1}, proves that our
schedule guarantees interference-freeness for all links.

\textbf{$\Leftarrow$:}  if no solution to the Partition problem
exists, this implies that
 for every partition of ${\cal I}$ into two subsets, the
sum of one set is greater than $N/2$. Assume we could still find a
schedule with only two slots. Since the receivers $r_{n+1}$ and
$r_{n+2}$ are at the same position,
 they have to be assigned to different slots.
Assume $r_{n+1}$ is assigned to the set with sum greater than $N/2$,
then the SINR at $r_{n+1}$ is below $\sigma$, which prevents the
correct reception of the signal from $s_{n+1}$,
 this causes contradiction.
\end{proof}

\subsection*{Proof of Lemma~\ref{l_sufficient}}

\begin{proof}
Consider any link $a=\left(  u,v\right)$. The wanted signal strength
is
\[
c\left\Vert a\right\Vert ^{\beta}\cdot\eta\left\Vert a\right\Vert ^{-{\kappa}%
}=c\eta\left\Vert a\right\Vert ^{\beta-{\kappa}}\geq c\eta R^{\beta-{\kappa}}.
\]
Consider any link $a^{\prime}=\left(  u^{\prime},v^{\prime}\right)
$ in $I$ other than $a$. We have $\left\Vert u^{\prime}u\right\Vert
\geq K\ell$. Therefore,
\[
\left\Vert u^{\prime}v\right\Vert \geq\left\Vert u^{\prime}u\right\Vert
-\left\Vert uv\right\Vert \geq K\ell-R=\left(  K/\sqrt{2}-1\right)  R.
\]
The total interference to $a$ from all other links in $I$ is at most%
\begin{align*}
&  \sum_{\left(  x,y\right)  \in\mathbb{Z}^{2}\setminus\left\{  \left(
0,0\right)  \right\}  }cR^{\beta}\cdot\eta\left(  \sqrt{x^{2}+y^{2}}%
\cdot\left(  K/\sqrt{2}-1\right)  R\right)  ^{-{\kappa}}\newline\\
&  =c\eta R^{\beta-{\kappa}}\left(  K/\sqrt{2}-1\right)  ^{-{\kappa}}%
\newline\sum_{\left(  x,y\right)  \in\mathbb{Z}^{2}\setminus\left\{  \left(
0,0\right)  \right\}  }\left(  \sqrt{x^{2}+y^{2}}\right)  ^{-{\kappa}%
}\newline\\
&  \leq4c\eta R^{\beta-{\kappa}}\left(  K/\sqrt{2}-1\right)  ^{-{\kappa}%
}\newline\left(  \sum_{i=1}^{\infty}i^{-{\kappa}}+\sum_{x=1}^{\infty}%
\sum_{y=1}^{\infty}\left(  \sqrt{x^{2}+y^{2}}\right)  ^{-{\kappa}}\right)
\newline\\
&  \leq4c\eta R^{\beta-{\kappa}}\left(  K/\sqrt{2}-1\right)  ^{-{\kappa}%
}\left(  \frac{{\kappa}(1+2^{-\frac{{\kappa}}{2}})}{{\kappa}-1}+\frac
{\pi2^{-{\kappa}/2}}{2({\kappa}-2)}\right)  \\
&  =4\tau c\eta R^{\beta-{\kappa}}\left(  K/\sqrt{2}-1\right)
^{-{\kappa}}%
\newline,
\end{align*}
\newline where
\[
\tau=\frac{{\kappa}(1+2^{-\frac{{\kappa}}{2}})}{{\kappa}-1}+\frac
{\pi2^{-{{\kappa}}/{2}}}{2({\kappa}-2)}.
\]
Thus the SINR at the receiver \ of the link  is at least
\[
\frac{c\eta R^{\beta-{\kappa}}}{\xi+4\tau c\eta R^{\beta-{\kappa}}\left(  K/\sqrt{2}-1\right)
^{-{\kappa}}\newline}\geq{\sigma}%
\]
since
\[
 K/\sqrt{2}-1\geq\left({(4\tau)}^{-1}\left({{\sigma}^{-1}-\xi{(c\eta)}^{-1}R^{\kappa-\beta}}\right)\right)^{-1/\kappa}
\]
\end{proof}
\end{document}